\documentclass[11pt]{article} 
\usepackage{hfstyle}
\usepackage{xcolor}
\usepackage{soul}
\usepackage{graphicx,bbm}
\usepackage{cite}
\usepackage[right]{eurosym}
\usepackage[T1]{fontenc}
\usepackage{amsmath}

\usepackage{bibtopic,url}
\usepackage[pdftex,pdfstartview=Fit,pdfpagemode=None,colorlinks=true,
pageanchor=true,linktocpage=false,filecolor=blue,pdfnewwindow=true,
linkcolor=blue]{hyperref}

\usepackage{hfstyle}
 
\begin{document}
\title{Remembering Nino Boccara (1931--2018)}
\author{Henryk Fuk\'s
      \oneaddress{
         Department of Mathematics and Statistics,
 Brock University,\\
     St. Catharines, Ontario, Canada  \\
         \email{hfuks@brocku.ca}
       }
   }

%
\Abstract{
In commemoration of the fifth anniversary since Nino Boccara's departure, this article
offers some personal recollections and provides insight into his life and accomplishments. Detailed bibliography of his works is included together with commentary highlighting his major achievements.
}
\maketitle

\section{Brief biography}
%
%
%

%
%

Nino (Samson) Boccara was born in Tunis on May 30, 1931,
but  when he was still a child, his family  moved to France.
He attended one of the most reputable schools in Paris, the École alsacienne, and in 1952 entered the highly competitive École supérieure de physique et de chimie industrielles de la ville de Paris (ESPCI Paris), one of the top engineering schools in France, today famous for having five Nobel prize winners  among its former professors and alumni.  
He graduated from ESPCI in 1956 as a member of the 71st graduating class with the diploma of engineer and then
 continued as a graduate student at the University of Paris (Sorbonne) under the direction of 
Jean Laval (1900--1980), professor of theoretical physics at the Collège de France from 1950 to 1970.

 In December 1961, he defended the doctoral thesis entitled \emph{Étude de l'effet Compton et des ondes élastiques dans les cristaux de sylvine}
 (\emph{Study of the Compton effect and elastic waves in sylvite crystals})
  receiving the degree of ``Docteur ès sciences physiques'' (PhD).
After the doctorate he worked for the next two years at    the
Laboratory of Theoretical Physics of Collège de France,
and  in 1964 secured a position at the Centre d'Etudes Nucléaires de Saclay,
research centre belonging to the Atomic Energy Commision (CEA), today known as
CEA Paris-Saclay. At first, he worked in the unit called Solid State Physics and Magnetic Resonace
Service, and later, after reorganization which happened around 1989/1990, in DRECAM/SPEC (Research Division on Atoms, Molecules and Condensed Matter/Service of Condensed Matter Physics). He remained affiliated with Saclay until retirement.

In addition to his position at Saclay, he  held appointments at several other institutions. 
In 1977 he became a professor of mathematics at ESPCI, where he taught various courses in mathematics for many years,
continuing well into late 1990's. He held the title of \emph{directeur de recherche} (senior scientist) at the CNRS,
Centre national de la recherche scientifique (National Centre for Scientific Research).
Furthermore,  for a number of years he served as a director of Centre de Physique at Les Houches,
where  throughout 1980's and in early 90's he organized a series of workshops
and edited  proceedings books \cite{ boccara1981symmetries,boccara85physics,
leger87polycyclic,maret1986biophysical,mlb87,lelay89semiconductor,
richard1988elementary,jullien1988universalities,manneville1989cellular,
polian1989simple,cb90,bgmp93}.
In 1989, he joined the Physics Department of the University of Illinois at Chicago
as a full professor, a position which he also held until retirement. He had some kind of arrangement with both UIC and Saclay which allowed him to spend half of the year in Paris and the other half in  Chicago. 

He retired from UIC in 2002, and around that time  also  from other positions he
held in France. He continued research and publishing activities well beyond this date,  with the  last research paper  published in 2010.
Nino Boccara died in Paris on December 16,  2018 at the age of 87.
\begin{figure}
\begin{center}
\includegraphics[width=12cm]{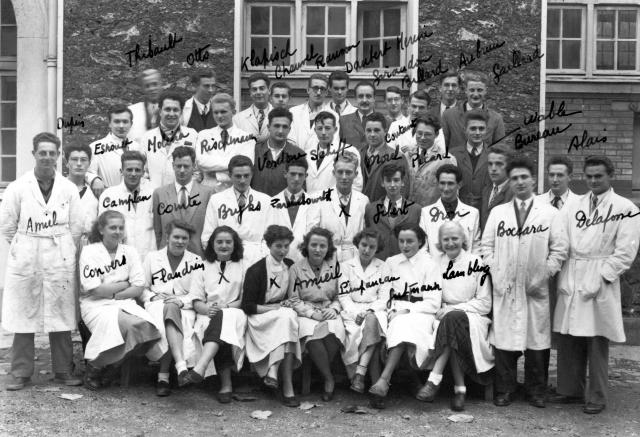}\\[1em]
\includegraphics[width=10cm]{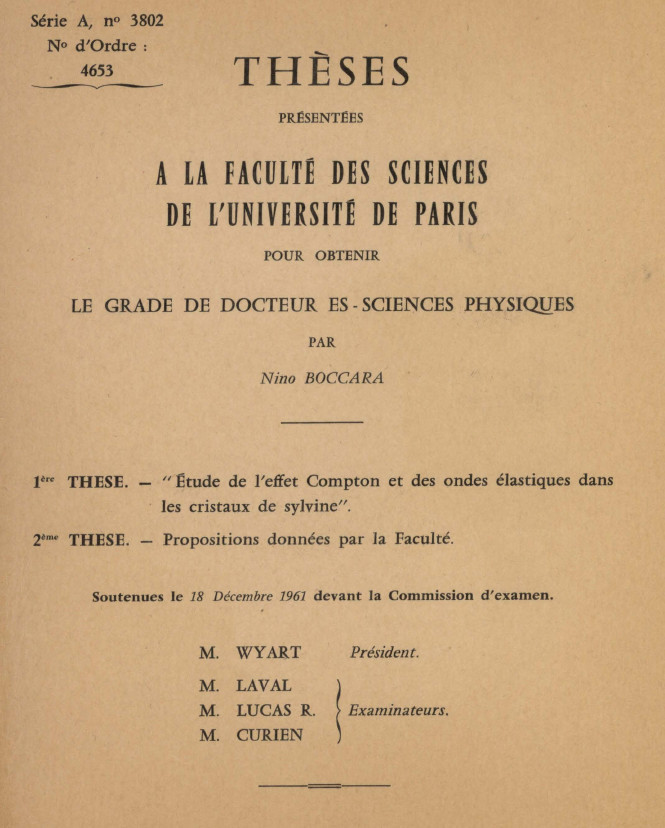}
\caption{Top: the 71st graduating class of ESPCI, 1957 (from \url{https://www.espci.org/fr/trombi/medium/71.jpg}). N.B. in the first row standing, third from the right. Bottom: Front page of the PhD thesis of N.B. from 1961. Reproduced from \url{https://inis.iaea.org/collection/NCLCollectionStore/_Public/54/034/54034400.pdf}.
}
\end{center}
\end{figure}

\section{Personal recollections}

I met Nino Boccara for the first time relatively late in his career, in 1993, when I started 
my graduate studies at the University of Illinois at Chicago.
After passing the preliminary exam,  I was looking for an advisor,
and then I read the departamental brochure describing research of all physics professors.
In this brochure, his research was described under the heading  ,,Discrete Dynamical Systems'' and the text below the heading was as follows:
\begin{quotation}
Most models in population dynamics are usually
formulated in terms of differential equations or random processes.
These models cannot take into account the local character of the interactions, which is, on
the contrary, included in automata network models.
In collaboration with graduate students at UIC and other collaborators in France,
Professor Boccara has build up various models in epidemiology and ecology. More recently, he has submitted a report to the Illinois
Criminal Justice Authority on street gang growth control.
\end{quotation}

This was so very different than the research one would expect to find at the physics
department that I became very intrigued and decided to talk to him.  Shortly afterwards,
after two or three long meetings,  he officially became my advisor. For the next 
few years I worked with him on various aspects of the
theory and applications of cellular automata. We continued collaboration
also after  my PhD and  published a number of joint papers, with the last one in 2006.

Nino Boccara taught at UIC various physics courses at both undergraduate and graduate 
level. In the Fall 1994, I took his Statistical Mechanics course, PHYS 561,
which I enjoyed tremendously, although I remember it as being very hard, especially the final exam. He incorporated cellular automata (CA)
into the course and showed us how some methods of statistical physics
could be used to analyze dynamics of CA. This is when I learned about the generalization of the mean-field approximation known as the local structure theory, developed by Howard Gutowitz in late 80's.

When I came to UIC, he had several graduate students. 
His first student, Kyeong Taik Cheong, already defended his PhD thesis in 1992,
but Qi Zeng, Mark Oram and  Servie Geurten either continued or arrived during my stay.
Franco Bagnoli also started his PhD under Nino's direction at approximately the same time, 
at University of Paris VI, where Nino had one of his multiple affiliations. 

Nino was giving his students a lot of freedom to choose research topics
and did not really press too hard for results. His students, therefore,
had to be able to work independently and needed a measure of self-discipline
in order to make progress in thesis research. Because he was traveling a lot and spending time  at different institutions, this also meant that much of the interaction with him had to be done remotely. Here is how Franco Bagnoli
remembers working with Nino:

\begin{quote} 
I do not remember exactly when I met Nino for the first time, probably at
the workshop ``Complexity and Evolution'' in Les 
Houches\footnote{Workshop which took place at Les Houches in March 6-15, 1990.}
in 1990. We also met
on other occasions, likely in Turin. Anyhow, he was already one of the ``big'' names
in the field of cellular automata, the subject of my master thesis (1989), and 
I was familiar with many of his papers. In 1991, after having spent 
some time in Geneva (Switzerland) working with Michel Droz, I was hired by the
University of Florence as a technician, the system manager of the
Department of Applied Mathematics, which curiously at that time had no
network, no servers and just a couple of PCs. 
I won the competition for
that position because I used to work in a computer company during my master
studies, and because the examination was organized by mathematicians, with
questions about functional analysis and linear algebra mixed with
computation.

After having built from scratch the network and the Internet services of
the department, I decided that after all I preferred doing physics rather
than informatics for a living, so I decided to get a PhD. I was hired
full-time (although with a lot of freedom in my schedule), and therefore
I looked for a PhD abroad, since at that time in Italy it was
not possible to pursue a PhD program without a grant (and thus without
abandoning my job). I contacted some researchers in France and Germany and
finally I got a response from Nino, who agreed to be the director
of my PhD.

At that time he was working partially in ESPCI in Paris and in the
University of Illinois at Chicago, but he also had some affiliation or
connection with the University Paris VI (Pierre et Marie Curie), and CEA
Paris-Saclay. So he asked me which of these institutions I preferred for my
PhD. Since ESPCI was mainly an engineering school, I chose Paris VI which was also  
the easiest to reach from Florence, being in the center of Paris.
I did not want to leave my job, so I had to use my free time and
days off. In the end we have never worked in Paris VI, we always met in Saclay
when I traveled to Paris, and a couple of times in Chicago. He also owned
a house in central Italy, in Magione, near Trasimeno lake, and we met there
once. For most of the time, we collaborated by email (well before the
existence of skype or other chat systems), and I discussed my thesis in
1997 in Saclay.

I have to say that relatively little of my thesis originated from the
collaboration with Nino. We published only one paper together, while I
collected about 10 articles for my PhD. Afterwards, we met from time to time
on various occasions, and he participated as a teacher in a school that I
co-organized in Turin, in 2000. So, although Nino was always a brilliant
mind, I profited more from  him before and after my PhD than during it,
essentially because we had few occasions to interact, he was always
travelling and my mind was split  between computer science and theoretical
physics. Anyhow, I finally managed to become a researcher (in 2001) and
then a professor of physics (with computer science as a hobby), so I can
say that Nino  helped me in reaching my goals.
\end{quote}

In the mid-90's computers and internet technology were rapidly evolving and Nino was
an enthusiastic adopter of all new technologies. He was a big fan of Macintosh
computers and had in his office both desktop and laptop versions of Mac. Since
I did not share his enthusiasm for Apple, he sometimes tried to convince me how
much better  Macs are compared to  PCs, and usually I had to concede. 
This  was also the time when  many other operating systems competed: VM, Vax, OS2, and many flavours of Unix.  Nino tried to stay on top
of all these things and we experimented with running cellular automata
simulations on various servers which Physics Department had.
 I should add here that Nino was also a serious Mathematica user, incorporated symbolic algebra into his courses and later wrote a book on Mathematica programming \cite{boccara2007essentials}.

He possessed tremendous and very  broad knowledge of diverse areas of  mathematics.
For his students at ESPCI he wrote a series of textbooks ,,Mathematics For Engineers'', for 
courses in probability,  integration, analytic functions and theory of distributions
\cite{boccara1995probabilites,boccara1995integration,boccara1996fonctions,boccara1997distributions}.
His most popular mathematics textbook  was {\em Functional analysis: an introduction for physicists},
first published in French \cite{boccara1984analyse} and then in English by
Elsevier \cite{boccara1990functional}.
Speaking of books, he authored
11 books, but has never became a skilled typist. A fellow graduate student at UIC
once laughingly remarked that Nino wrote all these books using one finger only\ldots

Although, as Franco Bagnoli remarked above, Nino was frequently traveling and 
his students sometimes had to be patient to meet him in person, when he was
around he always had time for students and colleagues. 
He was a man of great culture,  spoke fluently three languages, 
widely traveled, and had a great range of interests. I always admired his erudition and 
his French-Italian sense of humor. 
He is sorely missed by  everyone fortunate enough to have known him.

Below, I will discuss the works of Nino Boccara in more or less chronological order. I will  place 
somewhat more emphasis on the period after 1989, primarily because I am more familiar with his research during the later stage of his career.

\section{Statistical physics, 1960--1989}
It is very difficult to describe the vast research output of Nino
Boccara in one short note. I will, therefore, only describe some highlights here,
without any pretension of being exhaustive. For the rest, I refer the reader to the
the bibliography section included at the end. I believe that the bibliography is complete for the period 1989--2010, but for earlier years I probably missed some publications
in conference proceedings or in some less  known journals.

As already mentioned, his doctoral thesis was on Compton scattering, 
and for a while he continued this line of work 
\cite{nb1960c,nb1960b,nb1960a,WOS:A1962WS83000003,WOS:A1962WS83600009,Wos:A19639216a00006,Wos:A19639257a00007,1963SSCom...1..168B}. In mid 1960's
his interests shifted toward the theory of phase transitions. 
At the beginning, he studied phase transitions induced by deformation of crystals.
His most cited paper from that time, 
 published in Annals of Physics in 1968 \cite{1968AnPhy..47...40B},
 examines phase transitions accompanied by a change in crystal structure and demonstrates 
 how to use symmetry arguments (Landau’s theory) to determine all the
order parameters and the order of the phase transition. Various aspects of
second-order phase transitions in crystals occupied him for the next 
five years 
\cite{1968SSCom...6..211B,1969JPSJS..26..167B,1969PhLA...28..474B,1969PhLA...28..659B,1969SSCom...7..331B,1971PSSBR..43...11B,Wos:A1972m176400003,1972SSCom..11...39B,1973AnPhy..76...72B}. The exception is the paper on statistical
properties of focal conic textures in liquid crystals \cite{Wos:A1973q279800017},
published in 1973  jointly with P.-G. de Gennes and four other authors.
It has an interesting mathematical background:  the authors describe iterative filling of space with smectic material using the Apollonian packing of circles. 

In late 60's and early 70's, Nino  produced three books dealing with topics in  thermodynamics and statistical physics 
\cite{boccara1968principes,boccara1970physique,boccara1976symetries}.
The first of them, \emph{Les principes de la thermodynamique classique}, 
is especially worth mentioning because it is a unique book in the statistical  physics literature.
It presents the ``classical'' thermodynamics using the axiomatic approach starting from the concept of state. Fundamental thermodynamic quantities are  derived as functions of state with some
postulated properties, which are then shown to be uniquely defined. The book is as mathematically
rigorous as needed and yet it is also very concise, managing to cover all essential topics in only 160 pages. I~was pleasantly surprised that it has been  recently digitized under the FeniXX program\footnote{FeniXX réédition numérique,
 \url{www.ebooksfenixx.fr}
}, thus it is 
available in the electronic format from major online bookstores, including Amazon.

In approximately 1976,  Nino started studying more abstract solvable models 
exhibiting phase transitions and related phenomena. These included $n$-vector model \cite{1976PhLA...56..161B},
dilute magnetic models \cite{1979PhLA...70..347B,1981JAP....52.1728B},
spin glasses \cite{1981PhLA...86..181B,1982JAP....53.2192B}, as well as various
flavours of the Ising model \cite{1982JPhC...15.1381B,Wos:A1983rk79400002}. I believe this happened because of the
strong interest in spin glasses and dilute magnetic materials which was developing at that time in the physics community, and also because this was a subject of experimental investigations at Saclay in the solid state division. One of the most interesting and frequently cited papers
of this period is the paper from 1977 \cite{Wos:A1977cu51900006},
written jointly with two other authors, considering an assembly of asymmetric ellipsoids coupled by a constant infinite-range interactions. This is a model
related to the Potts model and it exhibits two phase transitions. The paper
demonstrates that the model can be solved exactly.

Almost all papers which followed,  from \cite{1982JPhC...15.1381B} to \cite{1989JPCM....1.5721B},
written in the period 1982--1989, deal with the Ising model and spin glass models.
One of the most important of these is \cite{1983PhLA...94..185B}, titled 
\emph{Dilute Ising models: a simple theory}. In this paper Nino proposed
a conceptually simple and elegant method to explain behaviour of diluted magnetic system models. The method had the simplicity of the mean field approximation but yielding much more
accurate results. It was based  on the theorem that
a function defined on a finite set can be represented by a polynomial.

The papers on Ising model and its various generalizations and modifications form a substantial body of work, with over 20 publications in the period 1982-1989.
They are mostly written in collaboration with A. Benyoussef 
(Faculté des sciences de Rabat, Morocco) and after 1986 also  with several other coauthors.
Some of these papers investigate very complicated models, for example, in the 
system considered in \cite{1986PhRvB..34.7775B}, the
 parameter space is six dimensional and the renormalization
 group transformation which the authors use has 
69 fixed points describing a large variety of critical behaviors.

\section{Complex systems, 1989-2010}
After obtaining the appointment at the University of Illinois at Chicago in 1989, Nino focused his research on cellular automata and models of complex systems. This shift marked a significant change in his research direction, as he did not publish any more papers in statistical physics \emph{per se}, apart from two papers on the Potts model on fractals  \cite{Wos:A1991fe35700004,Wos:A1992gw43300006} where he is listed as a last author. This suggests that Nino wanted to make a significant impact in a different field and was determined to move forward without looking back. 

He started with the study of phase transitions in probabilistic cellular automata 
\cite{PhysRevA.39.3094} and dynamics of ``defects'' in deterministic rules  \cite{bnr90,bnr91,br91}. In 1993, he published an important (in my opinion) paper
\cite{b93}
which is not as widely known as it should, in which he systematically studied 
surjective homomorphisms (factors) between elementary rules, focusing on
homomorphisms which are themselves CA rules. His hope, I believe, was to find
 conjugacies between rules, but conjugacy requires bijection and there are no
 non-trivial elementary rules which would be bijections, so he considered  the
 ``next best thing'', namely surjective rules.

\begin{figure}
\begin{center}
\includegraphics[width=14cm]{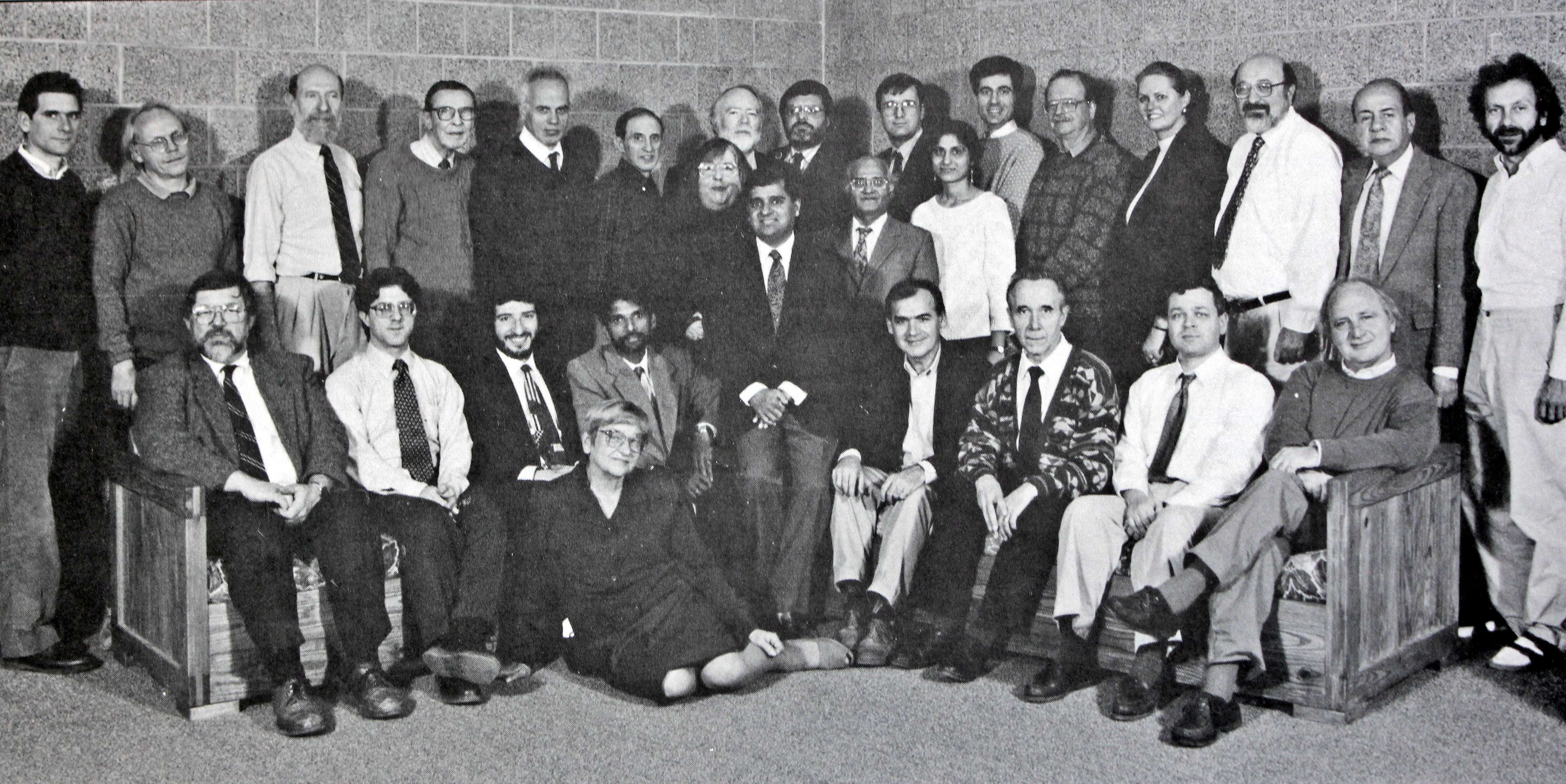}\\[1em]
\includegraphics[width=6cm]{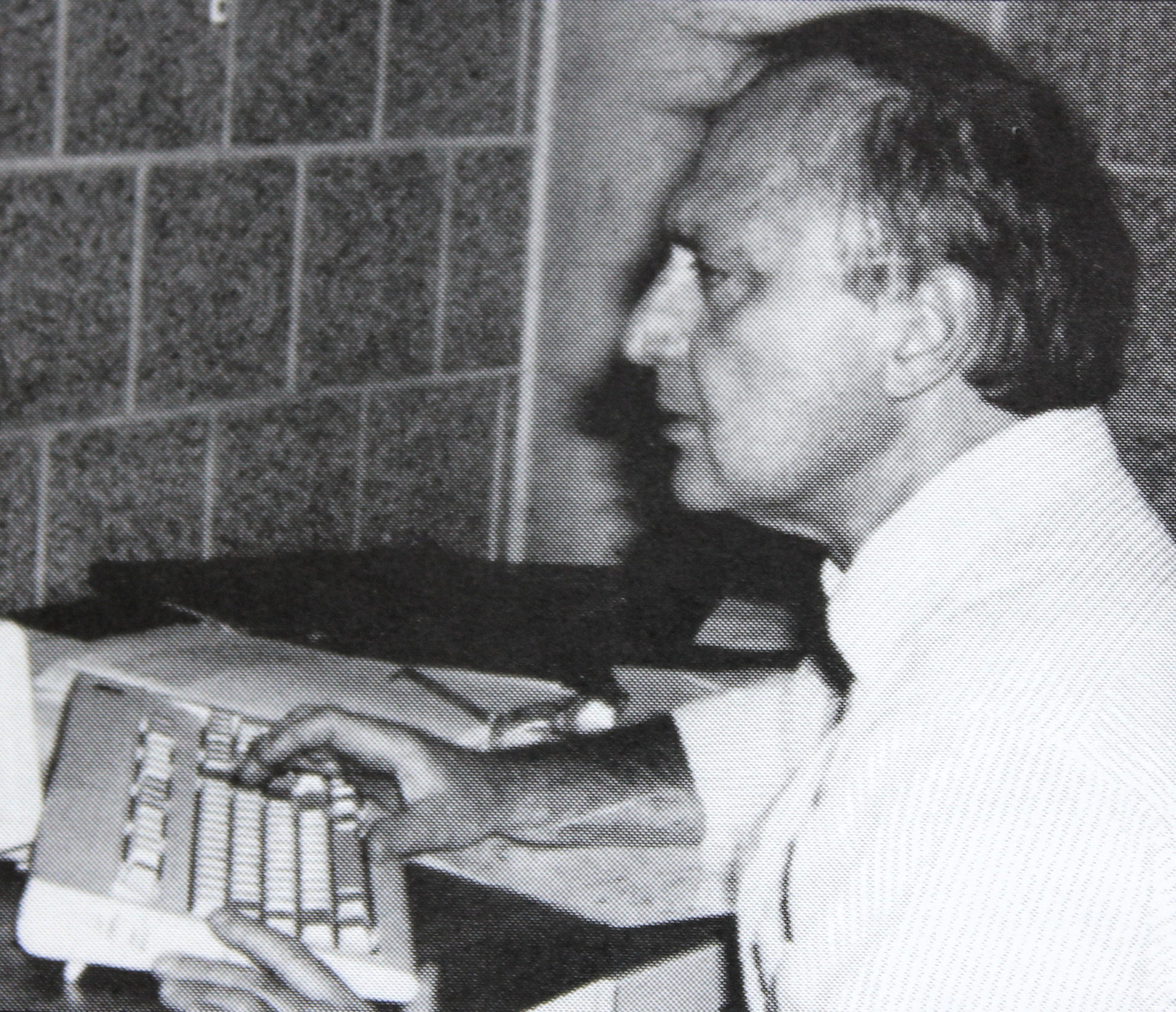}
\includegraphics[width=7cm]{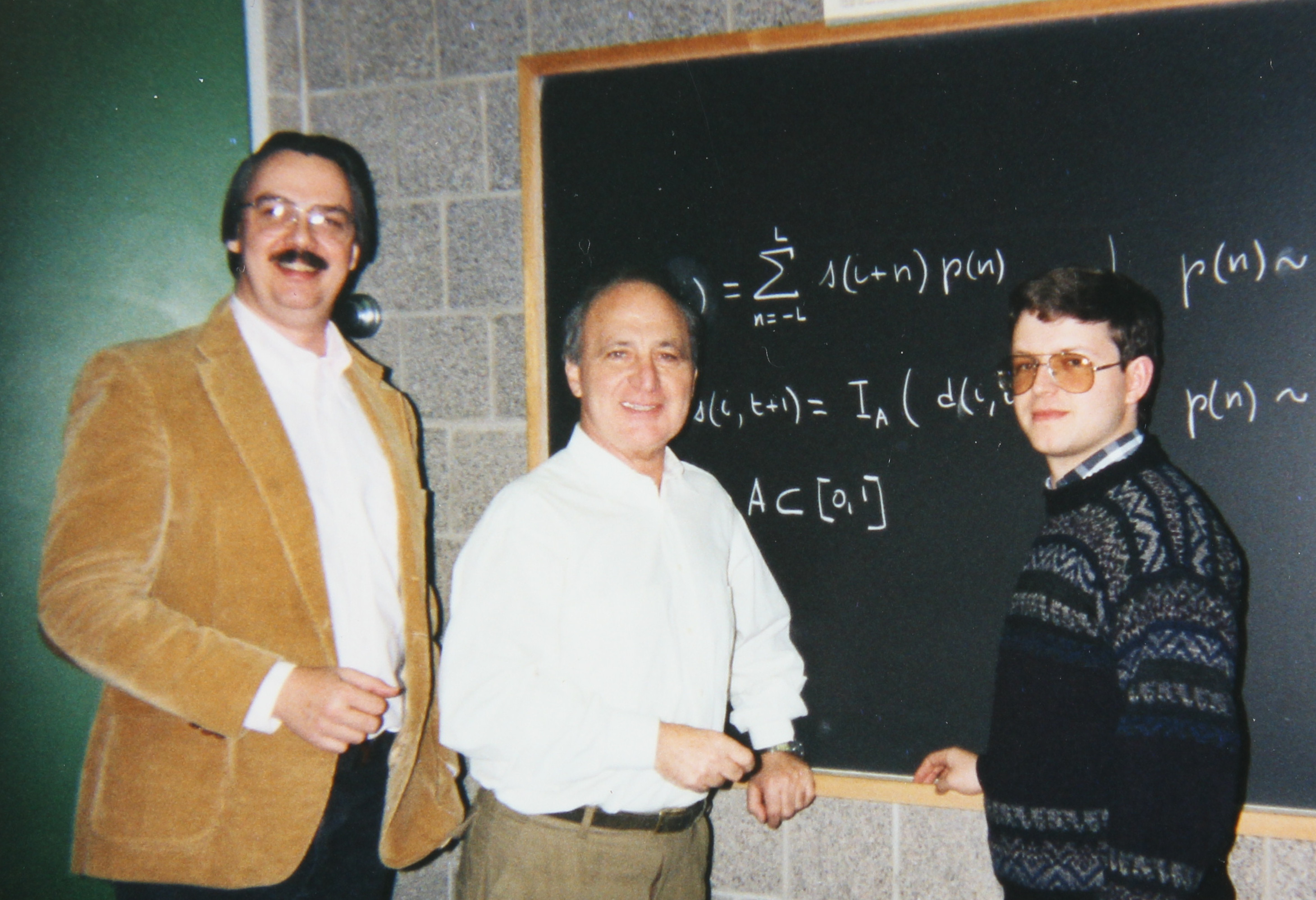}
\\[1em]
\includegraphics[width=7.5cm]{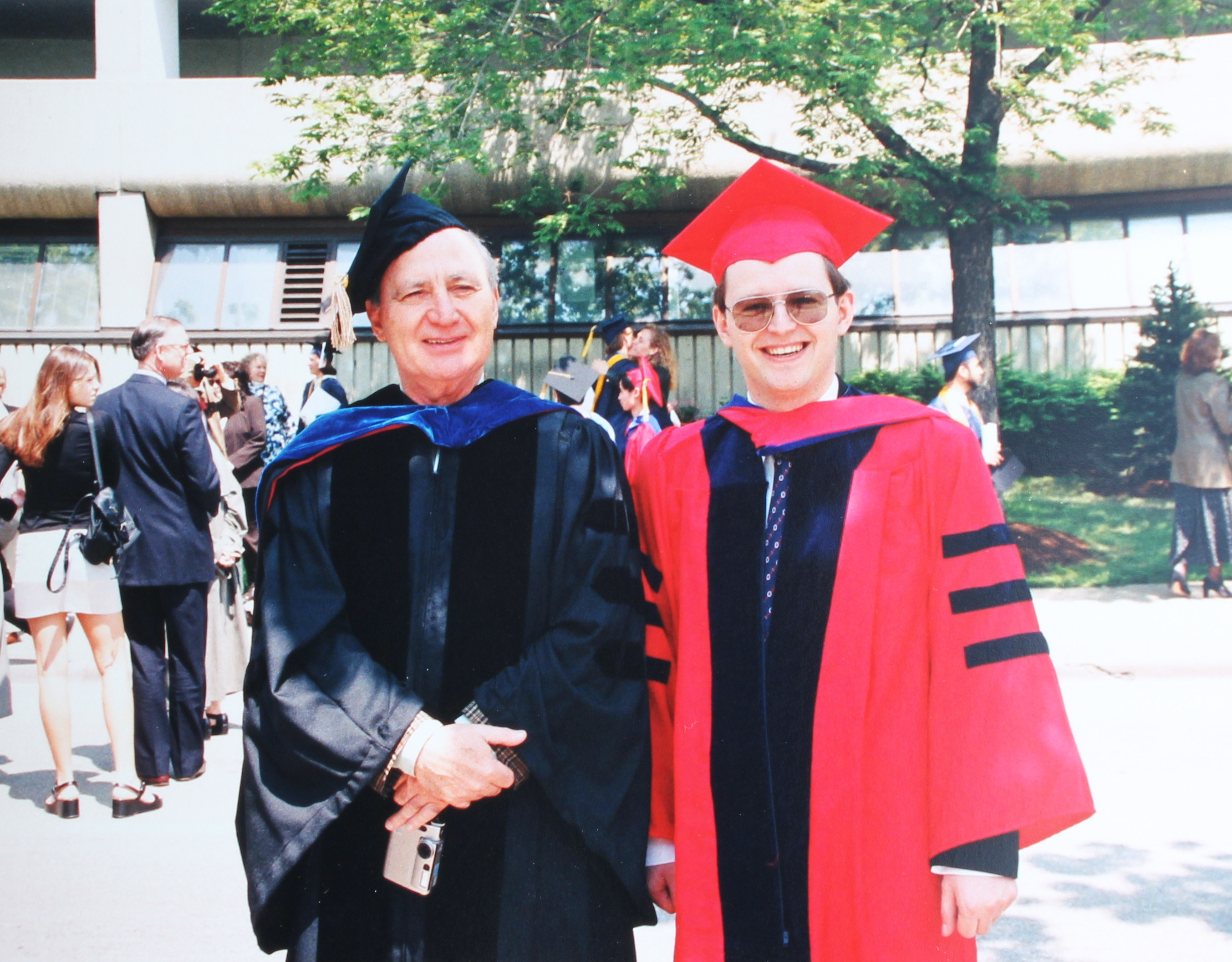}
\includegraphics[width=5cm]{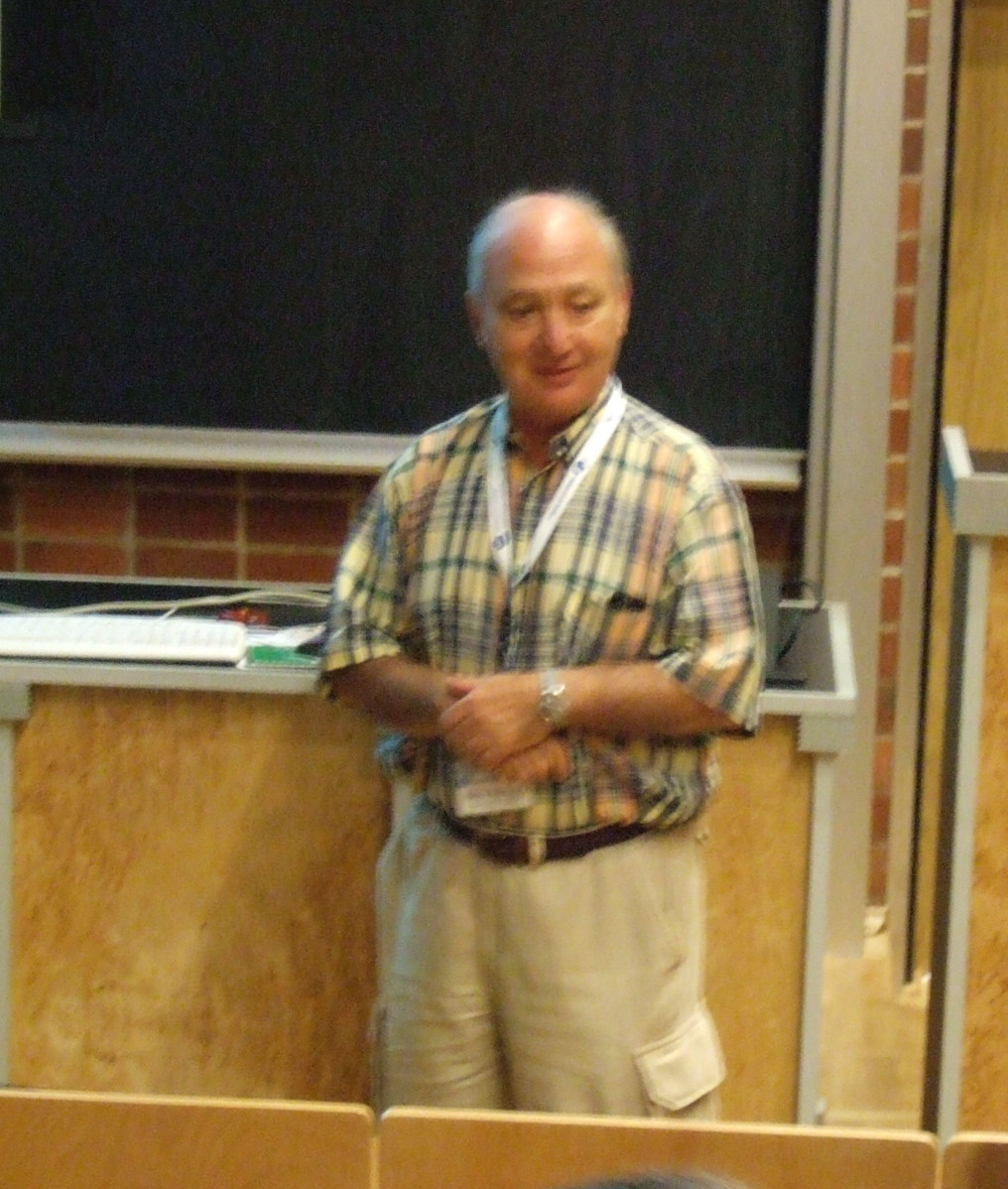}

\end{center}
\caption{
Top: UIC Physics Department, 1995, N.B. sitting first from the right.
Middle left: N.B. in front of his Mac computer in UIC office, mid 90's.
Middle right: with graduate students at UIC,  1995.
Bottom left: author with N.B., graduation ceremony at UIC, 1997.
Bottom right: lecture at Automata 2005, Gdańsk.
B\&W photos from the UIC Physic Dept. brochure (1995), the rest from author's collection. 
}
\end{figure}

 Around 1992 he came out with the idea of \emph{site exchange cellular automata},
 that is, cellular automata where site values of selected sites are repeatedly exchanged
 between iterations. The selection is usually random, but the sites selected for exchange
 can be either close neighbours (short range exchange) or completely uncorrelated (long range exchanges). He wanted to see how frequent these exchanges need to be to bring the cellular automaton to its mean-field or near mean-field regime. This idea appeared for the first time in the paper written with his first graduate student at UIC, K. Cheong
 \cite{bc92}. The cellular automaton used in the paper is a model of the epidemic
 process on a two-dimensional lattice. One of the main findings of this study was that when the rate of site exchanges tends to infinity the correlations are completely destroyed and the time evolution of the epidemic is then correctly predicted by the mean-field approximation.

The results of \cite{bc92} were quickly extended to a more general SIS epidemic model
\cite{bc93}, in which a transcritical bifurcation similar to a second-order phase transition
has been found. The third paper in the series \cite{bco94} added births and deaths to
the  epidemic model. This time the model revealed oscillatory behaviour of the
densities of  susceptible and infective individuals  as a function of time through a Hopf-type bifurcation parametrized by the rate of site exchanges.  A similar phenomenon appears in a version of the predator-prey 
model studied in \cite{brr94b}. The papers \cite{bc92,bc93,bco94,brr94b}
belong to the most frequently cited works of Nino Boccara among his papers
on CA and complex systems, and they continue to attract interest even now.
Later he co-authored another paper on population models investigating the spread of rabies among foxes  \cite{bbcez99}.

One of the problems which occupied Nino in early 90's was the question of existence of
collective global behaviour in cellular automata. Following the work of H. Chaté and P. Manneville
it was known that in ``classic'' cellular automata in dimensions lower than four
such behaviour is not possible. In \cite{bco94} (published in 1994 but based on results obtained in 1993 or earlier) global oscillations have been observed
in two dimensions, but this was because the range of interactions was made
effectively infinite by the site exchange process. Nino wondered if such 
increase of range could be made explicit by considering a weighted totalistic CA
in which the weight depends on the distance from the central site $r$ as
some function decreasing with $r$, for example, decreasing exponentially or as a power law.
He gave this problem to Servie Geurten, a newly arrived graduate student from Holland.
I~also started to work with Nino as my graduate supervisor almost at the same time (perhaps a few months later), thus  we became good friends with Servie. Unfortunately, shortly afterwards a tragedy struck: on December 30, 1993 Servie died in a car accident.
We felt devastated but I took over Servie's notes and we continued working on this problem for a while. Although it eventually drifted into a different direction then originally planned, we published a paper with Servie's name listed as a coauthor reporting the results. Due to editors sitting on it for  a very long  long time, it did not appear until 1997 \cite{bfg94}. 

After the famous paper of K. Nagel and M. Schreckenberg \emph{A cellular automaton model for freeway traffic} appeared in 1992, interest in CA traffic models was growing rapidly.
Nino became interested in this as well, and gave me the paper of Nagel and Schreckenberg
to read. We started to look into various aspects of traffic modelling and quickly
realized that although  some binary CA which conserve the sum of states can be used as 
simple models of traffic flow, there are many others which are interesting on their own right and these should be studied further. Nino coined the name ``number-conserving cellular automata''
for them, and we published several papers investigating their properties \cite{paper5,paper8,paper19,paper12,paper28}, the last three being the result of continued collaboration after I left UIC with PhD in 1997.
He later generalized the idea of number-conserving CA to monotone and eventually-conserving
CA \cite{MOREIRA2004285,bocc2007a}.

Although I remarked earlier  that after 1989 Nino have not published much in
the ``classical'' statistical physics, he still used various methods
of statistical physics to study dynamics of cellular automata.
For example, the 1993 paper \cite{obs93}, written with G. Ódor and G. Szabó, 
investigates the effect of mixing on one-dimensional probabilistic cellular automaton with totalistic rule, and it uses multiple-point-correlation approximation.
The authors found the  tricritical point for this system and estimated the critical exponent $\beta$. The paper with F. Bagnoli and P. Palmerini \cite{bbp98} from 1998 studies a different 
probabilistic rule, namely totalistic  cellular automaton having two absorbing states.
The automaton is found to exhibit  the second-order phase transition with the same critical behavior as the directed percolation model, in qualitative agreement with mean-field
predictions.

In the last five years of his active career Nino studied models
of the spread of globalized culture and opinion formation
\cite{bocc2005,bocc2007,bocc2008,WOS:000263880100006,WOS:000271586900002,WOS:000278088800005,WOS:000275469100001}. He was deeply concerned
about what American sociologist George Ritzer called \emph{the McDonaldization of society},
and wondered what could be done to keep alive  diverse local cultures and languages.
In \cite{bocc2007} he proposed a  model for the spread of a globalized culture in a population of individuals
located at the vertices of a scale-free social network. This was a model with synchronous update, thus one could call it a generalized  CA model. His conclusions were not very optimistic, but nevertheless recommended some concrete solutions:
\begin{quote}
In essence, this model shows that to resist globalization and keep alive a
diverse local culture, role models are essential, creations of as many links
as possible between individuals sharing the same local culture has to be
encouraged, and the society as a whole must make every effort to protect
and promote its own culture even at the cost of not fully respecting the rules
of free trade.
\end{quote}

The last paper of Nino Boccara was published in 2010 \cite{WOS:000275469100001}, but
I believe his last large project was the second edition of the book \emph{Modeling Complex Systems} \cite{compsys2003}. 
Originally published in 2003 by Springer, the book served as a textbook with the purpose of instructing readers in the construction of models of complex systems and equipping them with mathematical tools needed to analyze their dynamics. It was quite a success
and received glowing reviews in \emph{Physics Today}, \emph{Journal of Statistical Physics} and several other leading physics journals and magazines.
In 2009, Nino decided to update it and prepared the second edition, adding many new exercises with solution, extensively revising references and footnotes, and adding many features improving its  usability as a textbook.
The book includes discussion of Nino's own work, such as site-exchange, number-conserving 
and generalized cellular automata. It remains in print and I am sure it will
serve as a testament of his legacy  for many years to come.
\vskip 2em
\textbf{Acknowledgements}
I wish to thank Nazim Fatès and Anna Lawniczak for encouragement to write this 
article, Franco Bagnoli for contributing his recollections and
M. Catherine Kounelis for information about N.B. appointment at ESPCI. I am also very grateful to  Nazim Fatès for reading and correcting the manuscript.

\small
\section{Bibliography of works of Nino Boccara}
\bibliographystyle{hfbib}

\providecommand{\href}[2]{#2}\begingroup\raggedright\begin{thebibliography}{10}

\bibitem{boccara1968principes}
N.~Boccara, {\em Les principes de la thermodynamique classique}.
\newblock Collection Sup. Le physicien. Presses Universitaires de France,
  Paris, 1968.

\bibitem{boccara1970physique}
N.~Boccara, {\em La physique des transitions}.
\newblock Presses Universitaires de France, Paris, 1970.

\bibitem{boccara1976symetries}
N.~Boccara, {\em Sym{\'e}tries bris{\'e}es: th{\'e}orie des transitions avec
  param{\`e}tre d'ordre}.
\newblock Actualit{\'e}s scientifiques et industrielles. Hermann, Paris, 1976.

\bibitem{boccara1984analyse}
N.~Boccara and J.~Martin, {\em Analyse fonctionnelle: une introduction pour
  physiciens}.
\newblock ABC lexical. Ellipses, Paris, 1984.

\bibitem{boccara1990functional}
N.~Boccara, {\em Functional Analysis: An Introduction for Physicists}.
\newblock Elsevier Science, 1990.

\bibitem{boccara1995probabilites}
N.~Boccara, {\em Probabilit{\'e}s}.
\newblock Math{\'e}matiques pour l'ing{\'e}nieur. Ellipses, Paris, 1995.

\bibitem{boccara1995integration}
N.~Boccara, {\em Int{\'e}gration}.
\newblock Math{\'e}matiques pour l'ing{\'e}nieur. Ellipses, Paris, 1995.

\bibitem{boccara1996fonctions}
N.~Boccara, {\em Fonctions analytiques}.
\newblock Math{\'e}matiques pour l'ing{\'e}nieur. Ellipses, Paris, 1996.

\bibitem{boccara1997distributions}
N.~Boccara, {\em Distributions}.
\newblock Math{\'e}matiques pour l'ing{\'e}nieur. Ellipses, Paris, 1997.

\bibitem{compsys2003}
N.~Boccara, {\em Modeling Complex Systems}.
\newblock Springer, New York, 2003, 2010.

\bibitem{boccara2007essentials}
N.~Boccara, {\em Essentials of Mathematica: With Applications to Mathematics
  and Physics}.
\newblock Springer, New York, 2007.

\end{thebibliography}\endgroup


\providecommand{\href}[2]{#2}\begingroup\raggedright\begin{thebibliography}{10}

\bibitem{boccara1981symmetries}
N.~Boccara, {\em Symmetries and Broken Symmetries in Condensed Matter Physics:
  Proceedings of the Colloque Pierre Curie Held at the Ecole Sup{\'e}rieure de
  Physique Et de Chimie Industrielles de la Ville de Paris, Paris, September
  1980}.
\newblock IDSET, 1981.

\bibitem{boccara85physics}
N.~Boccara and M.~Daoud, {\em Physics of Finely Divided Matter: Proceedings of
  the Winter School, Les Houches, France, March 25--April 5, 1985}.
\newblock Springer Proceedings in Physics. Springer Berlin Heidelberg, 1985.

\bibitem{leger87polycyclic}
A.~L{\'e}ger, L.~D'Hendecourt, and N.~Boccara, {\em Polycyclic Aromatic
  Hydrocarbons and Astrophysics}.
\newblock Nato Science Series C:. Springer Netherlands, 1987.

\bibitem{maret1986biophysical}
G.~Maret, J.~Kiepenheuer, and N.~Boccara, {\em Biophysical Effects of Steady
  Magnetic Fields: Proceedings of the Workshop, Les Houches, France February
  26--March 5, 1986}.
\newblock Springer Proceedings in Physics. Springer Berlin Heidelberg, 1986.

\bibitem{mlb87}
J.~Meunier, D.~Langevin, and N.~Boccara, eds., {\em Physics of Amphiphilic
  Layers}.
\newblock Springer Berlin Heidelberg, 1987.

\bibitem{lelay89semiconductor}
G.~LeLay, J.~Derrien, and N.~Boccara, {\em Semiconductor Interfaces: Formation
  and Properties: Proceedings of the Workkshop, Les Houches, France February
  24--March 6, 1987}.
\newblock Springer Proceedings in Physics. Springer Berlin Heidelberg, 1987.

\bibitem{richard1988elementary}
J.~Richard, E.~Aslanides, and N.~Boccara, {\em The Elementary Structure of
  Matter: Proceedings of the Workshop, Les Houches, France, March 24--April 2,
  1987}.
\newblock Springer Proceedings in Physics. Springer Berlin Heidelberg, 1988.

\bibitem{jullien1988universalities}
R.~Jullien, L.~Peliti, R.~Rammal, and N.~Boccara, {\em Universalities in
  Condensed Matter: Proceedings of the Workshop, Les Houches, France, March
  15--25, 1988}.
\newblock Proceedings in Physics Series. Springer-Verlag, 1988.

\bibitem{manneville1989cellular}
P.~Manneville, N.~Boccara, G.~Vichniac, and R.~Bidaux, {\em Cellular Automata
  and Modeling of Complex Physical Systems: Proceedings of the Winter School,
  Les Houches, France, February 21--28, 1989}.
\newblock Springer Proceedings in Physics. Springer Berlin Heidelberg, 1989.

\bibitem{polian1989simple}
A.~Polian, P.~Loubeyre, N.~Boccara, and N.~A. T. O. S.~A. Division, {\em Simple
  Molecular Systems at Very High Density}.
\newblock NATO ASI series: Physics. Springer US, 1989.

\bibitem{cb90}
R.~Conte and N.~Boccara, eds., {\em Partially Integrable Evolution Equations in
  Physics}.
\newblock Nato Science Series C. Springer, 1990.

\bibitem{bgmp93}
N.~Boccara, E.~Goles, S.~Martinez, and P.~Picco, eds., {\em Cellular Automata
  and Cooperative Phenomena, Proc. of a Workshop, Les Houches, France, June 22
  to July 2, 1992}.
\newblock Kluwer Academic Publishers, Dordrecht, 1993.

\end{thebibliography}\endgroup


\providecommand{\href}[2]{#2}\begingroup\raggedright\begin{thebibliography}{10}

\bibitem{nb1960c}
N.~Boccara, ``Champ de forces du milieu cristallin et matrice de {F}ourier des
  halogénures alcalins,'' {\em Comptes rendus hebdomadaires des séances de
  l'Académie des sciences} {\bf 251} (1960) 2505--2507.

\bibitem{nb1960b}
N.~Boccara, ``Charge effective des ions qui forment los cristaux d'halogénuros
  alcalins. dé doublement des fréquences des oscillations de la branche
  optique,'' {\em Comptes rendus hebdomadaires des séances de l'Académie des
  sciences} {\bf 251} (1960) 1485.

\bibitem{nb1960a}
N.~Boccara, ``Fréquences des oscillations optiques longitudinales et
  trancvorsales dans la sylvine à 140° {K},'' {\em Comptes rendus
  hebdomadaires des séances de l'Académie des sciences} {\bf 250} (1960)
  1025.

\bibitem{WOS:A19629185A00011}
N.~Boccara, ``Electrostatique - potentiel et champ dune distribution volumique
  de multipoles,'' {\em Comptes rendus hebdomadaires des séances de
  l'Académie des sciences} {\bf 254} (1962), no.~6, 1011.

\bibitem{WOS:A1962WS83000003}
N.~Boccara, ``La déformation des ions dans les cristaux d'halogénures
  alcalins - cas des oscillations principales,'' {\em Journal de Physique et le
  Radium} {\bf 23} (1962), no.~5, 287--290.

\bibitem{WOS:A1962WS83600009}
N.~Boccara, ``Polarisation des vibrations atomiques dans les cristaux,'' {\em
  Journal de Physique et le Radium} {\bf 23} (1962), no.~11, 921--924.

\bibitem{Wos:A19639216a00006}
N.~Boccara, ``Electrostatique - généralisation du theoreme d'{E}arnshaw,''
  {\em Comptes rendus hebdomadaires des séances de l'Académie des sciences}
  {\bf 256} (1963), no.~11, 2317.

\bibitem{Wos:A19639257a00007}
N.~Boccara and A.~Zarembovitch, ``Physique des solides - étude de la
  polarisation des ondes elastiques de vitesse extremale se propageant dans les
  cristaux,'' {\em Comptes rendus hebdomadaires des séances de l'Académie des
  sciences} {\bf 257} (1963), no.~26, 4167.

\bibitem{1963SSCom...1..168B}
N.~{Boccara}, ``{Polarization of atomic vibrations in crystals},'' {\em Solid
  State Communications} {\bf 1} (Nov., 1963) 168--168.

\bibitem{Wos:A1964wu04100006}
N.~Boccara, ``Dynamique du reseau monoatomique unidimensionnel anharmonique,''
  {\em Journal De Physique} {\bf 25} (1964), no.~7, 748--750.

\bibitem{1964PhL.....8..311B}
N.~{Boccara}, ``{Sur la théorie de la dilatation thermique des cristaux à
  basse temperature},'' {\em Physics Letters} {\bf 8} (Mar., 1964) 311--312.

\bibitem{1967PhLA...24..679B}
N.~{Boccara}, ``{Sur la divergence de certaines grandeurs thermodynamiques au
  voisinage d'une transition du deuxi{\`e}me ordre},'' {\em Physics Letters A}
  {\bf 24} (June, 1967) 679--680.

\bibitem{1968AnPhy..47...40B}
N.~{Boccara}, ``{Second-order phase transitions characterized by a deformation
  of the unit cell},'' {\em Annals of Physics} {\bf 47} (Mar., 1968) 40--64.

\bibitem{1968SSCom...6..211B}
N.~{Boccara}, ``{Electric field influence on soft phonon modes frequencies in
  ferroelectrics},'' {\em Solid State Communications} {\bf 6} (Apr., 1968)
  211--213.

\bibitem{1969JPSJS..26..167B}
N.~{Boccara}, ``{Dynamical Properties of Crystals in External Fields
  Application to Phase Transitions},'' {\em Journal of the Physical Society of
  Japan Supplement} {\bf 26} (Jan., 1969) 167.

\bibitem{1969PhLA...28..474B}
N.~{Boccara}, ``{On the violation of Abelian groups in second order phase
  transitions},'' {\em Physics Letters A} {\bf 28} (Jan., 1969) 474--475.

\bibitem{1969PhLA...28..659B}
N.~{Boccara}, ``{Similar critical behaviour of certain singular quantities in
  the vecinity of a second-order transition line},'' {\em Physics Letters A}
  {\bf 28} (Feb., 1969) 659--660.

\bibitem{1969SSCom...7..331B}
N.~{Boccara}, ``{Exact thermodynamic relations and rigorous exponent equalities
  in the vicinity of a second-order transition line},'' {\em Solid State
  Communications} {\bf 7} (Feb., 1969) 331--333.

\bibitem{1971PSSBR..43...11B}
N.~{Boccara}, ``{Self-Consistent Theory of Displacive Phase Transitions in
  Magnetic Materials},'' {\em Physica Status Solidi B Basic Research} {\bf 43}
  (Jan., 1971) K11--K15.

\bibitem{Wos:A1972m176400003}
N.~Boccara, ``Phase transitions and critical phenomena in quantum systems,''
  {\em Berichte Der Bunsen-Gesellschaft Fur Physikalische Chemie} {\bf 76}
  (1972), no.~3-4, 193.

\bibitem{1972SSCom..11...39B}
N.~{Boccara}, ``{On the microscopic formulation of {L}andau theory of phase
  transitions},'' {\em Solid State Communications} {\bf 11} (July, 1972)
  39--40.

\bibitem{1973AnPhy..76...72B}
N.~{Boccara}, ``{Violation of rotational invariance and mesomorphic phase
  transitions characterized by an order parameter},'' {\em Annals of Physics}
  {\bf 76} (Mar., 1973) 72--79.

\bibitem{Wos:A1973q279800017}
R.~Bidaux, N.~Boccara, G.~Sarma, L.~D. Seze, P.-G. {{d}{e} Gennes}, and
  O.~Parodi, ``Statistical properties of focal conic textures in smectic
  liquid-crystals,'' {\em Journal De Physique} {\bf 34} (1973), no.~7,
  661--672.

\bibitem{Wos:A1974t659100003}
N.~Boccara and G.~Sarma, ``Does magnetic surface order exist?,'' {\em Journal
  De Physique Lettres} {\bf 35} (1974), no.~7-8, L95--L97.

\bibitem{1974PhLA...49..165B}
N.~{Boccara}, J.~P. {Carton}, and G.~{Sarma}, ``{On the theory of
  superconducting intercalated layer compounds},'' {\em Physics Letters A} {\bf
  49} (Aug., 1974) 165--167.

\bibitem{1976PhLA...56..161B}
N.~{Boccara}, ``{Critical temperature of a dilute d-dimensional n-vector
  model},'' {\em Physics Letters A} {\bf 56} (Mar., 1976) 161--162.

\bibitem{Wos:A1976bg86200002}
N.~Boccara, ``Some aspects of theory of 2nd-order transitions,'' {\em Molecular
  Crystals And Liquid Crystals} {\bf 32} (1976), no.~1-4, 1--4.

\bibitem{Wos:A1976bx68500008}
N.~Boccara, R.~Mejdani, and L.~Deseze, ``Lattice gas of axially-symmetric
  ellipsoids,'' {\em Molecular Crystals And Liquid Crystals} {\bf 35} (1976),
  no.~1-2, 91--108.

\bibitem{Wos:A1977cu51900006}
N.~Boccara, R.~Mejdani, and L.~Deseze, ``Solvable model exhibiting a 1st-order
  phase transition,'' {\em Journal de Physique} {\bf 38} (1977), no.~2,
  149--151.

\bibitem{Wos:A1977ea43400009}
N.~Boccara and P.~Maitte, ``Ultraviolet-spectra of acyclic cross-conjugated
  dienones,'' {\em Tetrahedron Letters} (1977), no.~46, 4031--4032.

\bibitem{1979PhLA...70..347B}
N.~{Boccara}, A.~{Ben Youssef}, and M.~{Hamedoun}, ``{Critical temperature of a
  dilute magnetic system with nearest-neighbour and next-nearest-neighbour
  interactions},'' {\em Physics Letters A} {\bf 70} (Mar., 1979) 347--348.

\bibitem{1981JAP....52.1728B}
N.~{Boccara}, ``{Abstract: Theoretical investigations of dilute magnetic
  systems},'' {\em Journal of Applied Physics} {\bf 52} (Mar., 1981) 1728.

\bibitem{1981PhLA...86..181B}
A.~{Benyoussef} and N.~{Boccara}, ``{Stability of spin-glass phases versus
  space dimensionality},'' {\em Physics Letters A} {\bf 86} (Nov., 1981)
  181--182.

\bibitem{1982JAP....53.2192B}
A.~{Benyoussef} and N.~{Boccara}, ``{Stability of spin-glass phases versus
  space dimensionality},'' {\em Journal of Applied Physics} {\bf 53} (Mar.,
  1982) 2192--2193.

\bibitem{1982JPhC...15.1381B}
A.~{Benyoussef} and N.~{Boccara}, ``{Phase diagrams of bond {I}sing models},''
  {\em Journal of Physics C Solid State Physics} {\bf 15} (Mar., 1982)
  1381--1389.

\bibitem{Wos:A1983rk79400002}
A.~Benyoussef and N.~Boccara, ``Phase-diagrams of random {I}sing models -
  simple and systematic successive-approximations,'' {\em Journal De Physique}
  {\bf 44} (1983), no.~10, 1143--1147.

\bibitem{1983PhLA...93..351B}
A.~{Benyoussef} and N.~{Boccara}, ``{Real space renormalization group
  investigation of three-dimensional {I}sing spin glasses},'' {\em Physics
  Letters A} {\bf 93} (Jan., 1983) 351--353.

\bibitem{1983PhLA...94..185B}
N.~{Boccara}, ``{Dilute {I}sing models: A simple theory},'' {\em Physics
  Letters A} {\bf 94} (Mar., 1983) 185--187.

\bibitem{1983JPhC...16.1901B}
A.~{Benyoussef} and N.~{Boccara}, ``{Existence of spin-glass phases for three-
  and four-dimensional {I}sing and Heisenberg models},'' {\em Journal of
  Physics C Solid State Physics} {\bf 16} (Apr., 1983) 1901--1918.

\bibitem{1984JPhC...17..285B}
A.~{Benyoussef} and N.~{Boccara}, ``{Real-space renormalisation-group analysis
  of a three-dimensional {I}sing spin glass},'' {\em Journal of Physics C Solid
  State Physics} {\bf 17} (Jan., 1984) 285--299.

\bibitem{Wos:A1983qc29900012}
A.~Benyoussef and N.~Boccara, ``Real space renormalization-group investigation
  of 3-dimensional {I}sing spin-glasses,'' {\em Physics Letters A} {\bf 93}
  (1983), no.~7, 351--353.

\bibitem{Wos:A1984rz60800016}
A.~Benyoussef and N.~Boccara, ``Real-space renormalisation-group analysis of a
  3-dimensional {I}sing spin-glass,'' {\em Journal Of Physics C-Solid State
  Physics} {\bf 17} (1984), no.~2, 285--299.

\bibitem{1984JAP....55.1667B}
A.~{Benyoussef} and N.~{Boccara}, ``{Lower critical dimensionality of the
  spin-glass phase transition},'' {\em Journal of Applied Physics} {\bf 55}
  (Mar., 1984) 1667--1668.

\bibitem{1984JAP....55.2419B}
A.~{Benyoussef} and N.~{Boccara}, ``{Random {I}sing models: Convergence of
  successive approximations towards exact results},'' {\em Journal of Applied
  Physics} {\bf 55} (Mar., 1984) 2419--2420.

\bibitem{1984JPhA...17L.547B}
N.~{Boccara}, ``{Dilute {I}sing model on fractal lattices},'' {\em Journal of
  Physics A Mathematical General} {\bf 17} (July, 1984) L547--L549.

\bibitem{1985JPhC...18.1899B}
N.~{Benayad}, A.~{Benyoussef}, and N.~{Boccara}, ``{The dilute spin-1 {I}sing
  model with crystal-field interactions},'' {\em Journal of Physics C Solid
  State Physics} {\bf 18} (Mar., 1985) 1899--1907.

\bibitem{1985JPhC...18.4275B}
A.~{Benyoussef}, N.~{Boccara}, and M.~{Saber}, ``{Dilute semi-infinite {I}sing
  model},'' {\em Journal of Physics C Solid State Physics} {\bf 18} (Aug.,
  1985) 4275--4289.

\bibitem{1986JPhC...19.1983B}
A.~{Benyoussef}, N.~{Boccara}, and M.~{Saber}, ``{Three-dimensional
  semi-infinite spin-1 {I}sing model interaction with crystal field},'' {\em
  Journal of Physics C Solid State Physics} {\bf 19} (Apr., 1986) 1983--1991.

\bibitem{1986JSP....45..113B}
R.~{Bidaux}, N.~{Boccara}, and G.~{Forg{\`a}cs}, ``{Three-spin interaction
  {I}sing model with a nondegenerate ground state at zero applied field},''
  {\em Journal of Statistical Physics} {\bf 45} (Oct., 1986) 113--134.

\bibitem{1986PhRvB..34.4881B}
R.~{Bidaux} and N.~{Boccara}, ``{Order of the phase transition in a
  three-dimensional {I}sing model with three-spin interactions},'' {\em Phys.
  Rev. B} {\bf 34} (Oct., 1986) 4881--4884.

\bibitem{1986PhRvB..34.7775B}
A.~{Benyoussef}, N.~{Boccara}, and M.~{El Bouziani}, ``{Real-space
  renormalization-group investigation of the three-dimensional semi-infinite
  Blume-Emery-Griffiths model},'' {\em Phys. Rev. B} {\bf 34} (Dec., 1986)
  7775--7783.

\bibitem{1987JPhC...20.2053B}
N.~{Benayad}, A.~{Benyoussef}, and N.~{Boccara}, ``{Re-entrant ferromagnetism
  in diluted two-dimensional {I}sing models with competitive nearest- and
  next-nearest-neighbour interactions},'' {\em Journal of Physics C Solid State
  Physics} {\bf 20} (May, 1987) 2053--2061.

\bibitem{1988JAP....63.4001B}
N.~{Benayad}, A.~{Benyoussef}, and N.~{Boccara}, ``{Reentrant ferromagnetism in
  a two-dimensional {I}sing model with random nearest-neighbor interactions
  (abstract)},'' {\em Journal of Applied Physics} {\bf 63} (Apr., 1988) 4001.

\bibitem{1988JPhC...21.5417B}
N.~{Benavad}, A.~{Benyoussef}, and N.~{Boccara}, ``{Re-entrant ferromagnetism
  in a two-dimensional {I}sing model with random nearest-neighbour
  interactions},'' {\em Journal of Physics C Solid State Physics} {\bf 21}
  (Nov., 1988) 5417--5425.

\bibitem{1988JPhC...21.5747B}
N.~{Benayad}, A.~{Benyoussef}, N.~{Boccara}, and A.~{El Kenz},
  ``{Renormalisation group recursion relations using the application of
  generalised Callen identities to the Ashkin-Teller model},'' {\em Journal of
  Physics C Solid State Physics} {\bf 21} (Dec., 1988) 5747--5756.

\bibitem{1989JPCM....1.5721B}
N.~{Boccara}, A.~{Elkenz}, and M.~{Saber}, ``{Mean-field theory of the spin-1
  {I}sing model with a random crystal field},'' {\em Journal of Physics
  Condensed Matter} {\bf 1} (Aug., 1989) 5721--5724.

\end{thebibliography}\endgroup


\providecommand{\href}[2]{#2}\begingroup\raggedright\begin{thebibliography}{10}

\bibitem{PhysRevA.39.3094}
R.~Bidaux, N.~Boccara, and H.~Chat\'e, ``Order of the transition versus space
  dimension in a family of cellular automata,'' {\em Phys. Rev. A} {\bf 39}
  (Mar, 1989) 3094--3105.

\bibitem{b89}
N.~Boccara, ``Scaling properties of a transformation defined on cellular
  automaton rules,'' {\em J. Phys. A: Math. Gen.} {\bf 22} (1989) L393--L396.

\bibitem{bnr90}
N.~Boccara, J.~Nasser, and M.~Roger, ``Annihilation of defects during the
  evolution of some one-dimensional class-3 deterministic cellular automata,''
  {\em Europhys. Lett.} {\bf 13} (1990), no.~6, 489--494.

\bibitem{br91}
N.~Boccara and M.~Roger, ``Block transformations of one-dimensional
  deterministic class-3 cellular automata,'' {\em J. Phys. A: Math. Gen.} {\bf
  24} (1991) 1849--1865.

\bibitem{Wos:A1991fe35700004}
A.~Bakchich, A.~Benyoussef, and N.~Boccara, ``The dilute {P}otts model on
  fractal lattices,'' {\em Journal Of Physics-Condensed Matter} {\bf 3} (Mar
  25, 1991) 1727--1739.

\bibitem{bnr91}
N.~Boccara, J.~Nasser, and M.~Roger, ``Particlelike structures and their
  interactions in spatiotemporal patterns generated by one-dimensional
  deterministic cellular-automaton rules,'' {\em Phys. Rev. A} {\bf 44} (1991),
  no.~2, 866--875.

\bibitem{Wos:A1992gw43300006}
A.~Bakchich, A.~Benyoussef, and N.~Boccara, ``Antiferromagnetic {P}otts model
  on {S}ierpinski carpets,'' {\em Journal De Physique I} {\bf 2} (Jan, 1992)
  41--54.

\bibitem{br92}
N.~Boccara and M.~Roger, ``Period-doubling route to chaos for a global variable
  of a probabilistic automata network,'' {\em J. Phys.A: Math. Gen.} {\bf 25}
  (1992) L1009--L1014.

\bibitem{bc92}
N.~Boccara and K.~Cheong, ``Automata network sir models for the spread of
  infectious diseases in populations of moving individuals,'' {\em J. Phys. A:
  Math. Gen.} {\bf 25} (1992) 2447--2461.

\bibitem{b93}
N.~Boccara, ``Transformations of one-dimensional cellular automaton rules by
  translation-invariant local surjective mappings,'' {\em Physica D} {\bf 68}
  (1993) 416--426.

\bibitem{br93}
N.~Boccara and M.~Roger, ``Site-exchange cellular automata,'' in {\em
  Instabilities and Nonequilibrium Structures IV}, E.~Tirapegui and W.~Zeller,
  eds., pp.~109--118.
\newblock Kluwer Academic Publishers, Dordrecht, 1993.

\bibitem{obs93}
G.~Ódor, N.~Boccara, and G.~Szabó, ``Phase transition study of a
  one-dimensional probabilistic site-exchange cellular automaton,'' {\em Phys.
  Rev. E} {\bf 48} (1993) 3168--3171.

\bibitem{bc93}
N.~Boccara and K.~Cheong, ``Critical behaviour of a probabilistic automata
  network {S}{I}{S} model for the spread of an infectious disease in a
  population of moving individuals,'' {\em J.Phys. A: Math. Gen.} {\bf 26}
  (1993) 3707--3717.

\bibitem{BLCP:CN004869436}
N.~Boccara, J.~Nasser, and M.~Roger, ``Ideal patterns and defects of some
  families of one-dimensional deterministic cellular automata,'' in {\em
  Complex dynamics, Centre de Physique Les Houches series}, R.~Livi, J.-P.
  Nadal, and N.~Packard, eds., pp.~113--124.
\newblock Nova Science Publishers, 1993.

\bibitem{bnr94a}
N.~Boccara, J.~Nasser, and M.~Roger, ``Critical behavior of a probabilistic
  local and nonlocal site-exchange cellular automaton,'' {\em International
  Journal of Modern Physics C} {\bf 05} (1994), no.~03, 537--545.

\bibitem{b94}
N.~Boccara, ``Automata network models of interacting populations,'' in {\em
  Cellular Automata, Dynamical Systems and Neural Networks}, E.~Goles and
  S.~Martinez, eds., pp.~23--78.
\newblock Kluwer, Dordrecht, 1994.

\bibitem{bnr94}
N.~Boccara, J.~Nasser, and M.~Roger, ``Critical behavior of a probabilistic
  local and nonlocal site-exchange cellular automaton,'' {\em Int. J. Mod.
  Phys. C} {\bf 5} (1994), no.~3, 537--545.

\bibitem{br94}
N.~Boccara and M.~Roger, ``Some properties of local and nonlocal site-exchange
  deterministic cellular automata,'' {\em Int. J. Mod. Phys. C} {\bf 5} (1994),
  no.~3, 581--588.

\bibitem{bco94}
N.~Boccara, K.~Cheong, and M.~Oram, ``Probabilistic automata network epidemic
  model with births and deaths exhibiting cyclic behaviour,'' {\em J. Phys. A:
  Math. Gen.} {\bf 27} (1994) 1585--1597.

\bibitem{brr94a}
N.~Boccara, O.~Roblin, and M.~Roger, ``Route to chaos for a global variable of
  two-dimensional ``game of life type'' automata network,'' {\em J.Phys. A:
  Math. Gen.} {\bf 27} (1994) 8039--8047.

\bibitem{brr94b}
N.~Boccara, O.~Roblin, and M.~Roger, ``An automata network predator-prey model
  with pursuit and evasion,'' {\em Phys. Rev. E} {\bf 50} (1994) 4531--4541.

\bibitem{crane1995dynamics}
J.~Crane and N.~Boccara, {\em The Dynamics of Street Gang Growth and Police
  Response: A Formal Model with Implications for Policy}.
\newblock Working paper. Institute of Government and Public Affairs, University
  of Illinois, 1995.

\bibitem{bfg94}
N.~Boccara, H.~Fuk{\'s}, and S.~Geurten, ``A new class of automata networks,''
  {\em Physica D} {\bf 103} (1997) 145--154,
  \href{http://arxiv.org/abs/chao-dyn/9610002}{{\tt
  chao-dyn/9610002}}.

\bibitem{paper2}
H.~Fuk{\'s} and N.~Boccara, ``Cellular automata models for diffusion of
  innovations,'' {\em unpublished report} (1997)
  \href{http://arxiv.org/abs/adap-org/9704002}{{\tt
  adap-org/9704002}}.

\bibitem{paper6}
N.~Boccara, H.~Fuk{\'s}, and Q.~Zeng, ``Car accidents and number of stopped
  cars due to road blockage on a one-lane highway,'' {\em J. Phys. A: Math.
  Gen.} {\bf 30} (1997) 3329--3332,
  \href{http://arxiv.org/abs/adap-org/9704001}{{\tt
  adap-org/9704001}}.

\bibitem{bbp98}
F.~Bagnoli, N.~Boccara, and P.~Palmerini, ``Phase transitions in a
  probabilistic cellular automaton with two absorbing states,''
  \href{http://arxiv.org/abs/cond-mat/9705171}{{\tt
  cond-mat/9705171}}. Lecture notes of the Summer School on Biotechnology
  held at Torino (Italy) in June 1996.

\bibitem{paper5}
H.~Fuk{\'s} and N.~Boccara, ``Generalized deterministic traffic rules,'' {\em
  Int. J. Mod. Phys. C} {\bf 9} (1998) 1--12,
  \href{http://arxiv.org/abs/adap-org/9705003}{{\tt
  adap-org/9705003}}.

\bibitem{paper7}
N.~Boccara and H.~Fuk{\'s}, ``Modeling diffusion of innovations with
  probabilistic cellular automata,'' in {\em Cellular Automata: A Parallel
  Model}, M.~Delorme and J.~Mazoyer, eds.
\newblock Kluwer Academic Publishers, Dordrecht, 1998.
\newblock \href{http://arxiv.org/abs/adap-org/9705004}{{\tt
  adap-org/9705004}}.

\bibitem{paper8}
N.~Boccara and H.~Fuk{\'s}, ``Cellular automaton rules conserving the number of
  active sites,'' {\em J. Phys. A: Math. Gen.} {\bf 31} (1998) 6007--6018,
  \href{http://arxiv.org/abs/adap-org/9712003}{{\tt
  adap-org/9712003}}.

\bibitem{bbcez99}
A.~Benyoussef, N.~Boccara, H.~Chakib, and H.~Ez-Zahraouy, ``Lattice
  three-species models of the spatial spread of rabies among foxes,'' {\em
  International Journal of Modern Physics C} {\bf 10} (1999), no.~06,
  1025--1038, \href{http://arxiv.org/abs/adap-org/9904005}{{\tt
  adap-org/9904005}}.

\bibitem{br99}
N.~Boccara and M.~Roger, ``Totalistic two-dimensional cellular automata
  exhibiting global periodic behavior,'' {\em International Journal of Modern
  Physics C} {\bf 10} (1999), no.~06, 1017--1024,
  \href{http://arxiv.org/abs/adap-org/9904002}{{\tt
  adap-org/9904002}}.

\bibitem{paper14}
N.~Boccara and H.~Fuk{\'s}, ``Critical behavior of a cellular automaton highway
  traffic model,'' {\em J. Phys. A: Math. Gen.} {\bf 33} (2000) 3407--3415,
  \href{http://arxiv.org/abs/cond-mat/9911039}{{\tt
  cond-mat/9911039}}.

\bibitem{bb2000}
R.~Bidaux and N.~Boccara, ``Correlated random walks with a finite memory
  range,'' {\em International Journal of Modern Physics C} {\bf 11} (2000),
  no.~05, 921--947, \href{http://arxiv.org/abs/adap-org/9904001}{{\tt
  adap-org/9904001}}.

\bibitem{bocc2000}
N.~Boccara, ``Automata network models in ecology and epidemiology,'' in {\em
  Dynamical Modeling in Biotechnology}, F.~Bagnoli and S.~Ruffo, eds.,
  pp.~73--96.
\newblock World Scientific, 2000.

\bibitem{bocc2001}
N.~Boccara, ``On the existence of a variational principle for deterministic
  cellular automaton models of highway traffic flow,'' {\em International
  Journal of Modern Physics C} {\bf 12} (2001), no.~02, 143--158,
  \href{http://arxiv.org/abs/cond-mat/0005244}{{\tt
  cond-mat/0005244}}.

\bibitem{paper19}
H.~Fuk{\'s} and N.~Boccara, ``Convergence to equilibrium in a class of
  interacting particle systems,'' {\em Phys. Rev. E} {\bf 64} (2001) 016117,
  \href{http://arxiv.org/abs/nlin.CG/0101037}{{\tt
  nlin.CG/0101037}}.

\bibitem{PhysRevE.63.046116}
F.~Bagnoli, N.~Boccara, and R.~Rechtman, ``Nature of phase transitions in a
  probabilistic cellular automaton with two absorbing states,'' {\em Phys. Rev.
  E} {\bf 63} (Mar, 2001) 046116,
  \href{http://arxiv.org/abs/cond-mat/0002361}{{\tt
  cond-mat/0002361}}.

\bibitem{paper12}
N.~Boccara and H.~Fuk{\'s}, ``Number-conserving cellular automaton rules,''
  {\em Fundamenta Informaticae} {\bf 52} (2002) 1--13,
  \href{http://arxiv.org/abs/adap-org/9905004}{{\tt
  adap-org/9905004}}.

\bibitem{MOREIRA2004285}
A.~Moreira, N.~Boccara, and E.~Goles, ``On conservative and monotone
  one-dimensional cellular automata and their particle representation,'' {\em
  Theoretical Computer Science} {\bf 325} (2004), no.~2, 285--316,
  \href{http://arxiv.org/abs/nlin/0306040}{{\tt nlin/0306040}}.

\bibitem{bocc2005}
N.~Boccara, ``Is it possible to protect cultural diversity? {A} mathematical
  model,'' {\em Il Pensiero Politico} {\bf 38} (2005) 454--464.

\bibitem{paper28}
N.~Boccara and H.~Fuk{\'s}, ``Motion representation of one-dimensional cellular
  automaton rules,'' {\em Int. J. Mod. Phys. C} {\bf 17} (2006) 1605--1611,
  \href{http://arxiv.org/abs/nlin.CG/0501043}{{\tt
  nlin.CG/0501043}}.

\bibitem{bocc2007a}
N.~Boccara, ``Eventually number-conserving cellular automata,'' {\em
  International Journal of Modern Physics C} {\bf 18} (2007), no.~01, 35--42,
  \href{http://arxiv.org/abs/cond-mat/0410563}{{\tt
  cond-mat/0410563}}.

\bibitem{bocc2007b}
N.~Boccara, ``Randomized cellular automata,'' {\em International Journal of
  Modern Physics C} {\bf 18} (2007), no.~08, 1303--1312,
  \href{http://arxiv.org/abs/nlin/0702046}{{\tt nlin/0702046}}.

\bibitem{bocc2007}
N.~Boccara, ``Is it possible to control the spread of a globalized culture?,''
  {\em J. of Cellular Automata} {\bf 2} (2007), no.~2, 111--120,
  \href{http://arxiv.org/abs/nlin/0611035}{{\tt nlin/0611035}}.

\bibitem{bocc2008}
N.~Boccara, ``Models of opinion formation: Influence of opinion leaders,'' {\em
  International Journal of Modern Physics C} {\bf 19} (2008), no.~01, 93--109,
  \href{http://arxiv.org/abs/0704.1790}{{\tt 0704.1790}}.

\bibitem{Boccara2009}
N.~Boccara, ``Phase transitions in cellular automata,'' in {\em Encyclopedia of
  Complexity and Systems Science}, R.~A. Meyers, ed., pp.~6771--6782.
\newblock Springer, New York, NY, 2009.

\bibitem{WOS:000263880100006}
N.~Boccara, ``A cellular automaton modeling the struggle to control the
  media,'' {\em International Journal Of Modern Physics C} {\bf 20} (2009),
  no.~2, 253--266.

\bibitem{WOS:000271586900002}
N.~Boccara, ``Maintenance and disappearance of minority languages: A cellular
  automaton model,'' {\em Journal of Cellular Automata} {\bf 4} (2009), no.~4,
  253--266.

\bibitem{WOS:000278088800005}
N.~Boccara, ``Evolution of extremist opinions in a population of interacting
  agents,'' {\em International Journal Of Modern Physics C} {\bf 21} (2010),
  no.~5, 617--628.

\bibitem{WOS:000275469100001}
N.~Boccara, ``Voters' fickleness: A mathematical model,'' {\em International
  Journal Of Modern Physics C} {\bf 21} (2010), no.~2, 149--158.

\end{thebibliography}\endgroup
\begin{btSect}[hfbib]{bocc-books}
 \subsection{Books}
 \btPrintAll
\end{btSect}

\begin{btSect}[hfbib]{bocc-edited}
 \subsection{Edited books}
 \btPrintAll
\end{btSect}

\begin{btSect}[hfbib]{bocc-articles-early-new}
 \subsection{Research papers 1960--1989}
 \btPrintAll
\end{btSect}

\begin{btSect}[hfbib]{bocc-articles}
 \subsection{Research papers 1989--2010}
 \btPrintAll
\end{btSect}

\end{document}